\documentclass[twocolumn,showpacs,preprintnumbers,amsmath,amssymb]{revtex4}

\usepackage{graphicx}
\usepackage{dcolumn}
\usepackage{bm}

\begin{document}

\preprint{APS/123-QED}

\title{
Pseudosymmetric bias and correct estimation of Coulomb/confinement energy \\
for unintentional quantum dot in channel of metal-oxide-semiconductor field-effect transistor
}

\author{K. Ono,$^1$ T. Tanamoto,$^2$ T. Ohguro$^2$}
\affiliation{
$^1$Low Temperature Physics Laboratory, RIKEN, 2-1 Hirosawa, Wako-shi, Saitama 351-0198, Japan\\
$^2$Corporate R\&D Center, Toshiba Corporation, 1 Komukai, Toshiba-cho, Saiwai-ku, Kawasaki-shi, Kanagawa 212-8582, Japan\\
}

\date{\today}

\begin{abstract}

We describe a measurement method that enables the correct estimation of the charging energy of an unintentional quantum dot (QD) in the channel of a metal-oxide-semiconductor field-effect transistor (MOSFET). If the channel has a dominant QD with a large charging energy and an array of stray QDs with much weaker charging, this method eliminates the additional voltage drops due to stray QDs by regarding the stray QDs as series resistors. We apply this method to a short-channel MOSFET and find that the charging energy of the dominant QD can indeed be smaller than the size of the Coulomb diamond.

\end{abstract}

\pacs{ 73.40.Qv, 73.23.-b, 73.43.Fj}

\maketitle 

The progress of quantum dot (QD) physics over the last two decades has resulted in a detailed understanding of quantum transport phenomena in QD devices \cite{QDreview}. Not only the charges but also the spins and nuclear spins in QDs have been electrically manipulated in QD devices based on III-V semiconductors \cite{Hanson}. Recently, quantum transport via a single trap site in silicon metal-oxide-semiconductor field-effect transistors (MOSFETs) has attracted considerable attention in terms of quantum information processing with electron spins \cite{Sellier, Y.Ono, Lansbergen, Tan, Pierre, Tabe, Fuechsle, Prati}. These trap sites in silicon, which essentially act as QDs, have various advantages over III-V QDs because the minor effects of nuclear spins enable the robust coherence of electron spins.

If a single trap site happens to exist in a short-channel MOSFET, single-electron transport via the trap site takes place in a subthreshold region of $V_{G}$ at a low temperature. The subthreshold region satisfies the following three conditions. 1) Direct transport or tunneling from the source to drain is negligible, $i.e.$, the channel is closed in the usual context of MOSFETs. 2) The energy level of the trap site, which is in the band gap of silicon, is comparable to the Fermi energies of the source and drain electrodes. 3) The channel is sufficiently short to allow tunneling between the trap site and the source or drain. Single-electron transport via a single dopant has been reported in the subthreshold region of a MOSFET \cite{Sellier, Y.Ono, Lansbergen, Tan, Pierre, Tabe, Fuechsle, Prati}. 

In a single-QD device, where a QD is weakly coupled to both the source and drain electrodes, and capacitively coupled to the gate electrode, the measurement of the drain current $I_{D}$ as a function of source voltage $V_{S}$ and gate voltage $V_{G}$ indicates a series of diamond-shaped regions where $I_{D}$ is strongly suppressed. The size of these Coulomb diamonds measured in $V_{S}$ is the energy of Coulomb charging and/or quantum confinement for the dot (hereafter we refer to this simply as the charging energy). Coulomb diamond measurement has been used as a powerful method for characterizing the single-electron transport in QD devices \cite{QDreview}. Clear Coulomb diamonds were observed and the charging energies of individual dopants were discussed on the basis of the size of the Coulomb diamonds \cite{Sellier, Y.Ono, Lansbergen, Tan, Pierre, Tabe, Fuechsle, Prati}.

\begin{figure}
\includegraphics[width=0.45\textwidth, keepaspectratio]{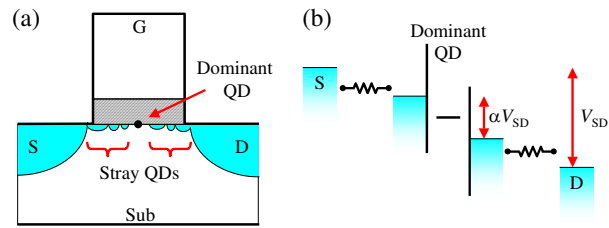}%
\caption{\label{fig.1} 
(a) Schematic of MOSFET channel with a single trap site (a dominant QD) marked as a black dot at the center of the channel, and some stray QDs caused by the potential fluctuation of the channel. (b) Schematic potential landscape for the channel. Stray QDs are modeled as series resistors that connect the source (drain) electrode with the tunnel junction to the dominant QD. 
}
\end{figure}

Under actual conditions, it is likely that, in addition to the single trap site (a dominant QD with large charging energy and limit $I_{D}$), a potential fluctuation near the dominant QD will act as a series of stray QDs that are distributed around the dominant QD (Fig. 1 (a)). These stray QDs will have much smaller charging energies. Thus, at temperatures where the thermal energy is comparable to or larger than the charging energies of the stray QDs, the array of stray QDs will behave as effective series resistors that are located between the dominant QD and the source (drain) electrode (Fig. 1 (b)). The series resistors cause additional drops of the source-drain voltage $V_{SD}$, as well as the usual voltage drop caused by the dominant QD. Owing to this series resistance, only a fraction $\alpha$  ($0 < \alpha < 1$) of $V_{SD}$ is applied to the dominant QD, and the size of the Coulomb diamond measured in $V_{S}$ will be larger than the actual charging energy of the dominant QD.
In this paper we describe a measurement method, a pseudosymmetric bias, which enables the correct estimation of $\alpha$. We apply this method to a short-channel MOSFET and evaluate $\alpha$.

Under the bias condition shown in Fig. 2 (a), the MOSFET channel is symmetrically biased with two identical electronics connected to the source and drain electrodes. This symmetric bias condition can be mimicked using more conventional measurement electronics as in Fig. 2 (b). Instead of using two identical electronics, suppose that we can shift the grounds of only the source and drain electronics, as in Fig. 2 (c), then the source and drain electrodes are biased with voltages of +$V_{S}$/2 and -$V_{S}$/2, respectively. A simpler setup is shown in Fig. 2 (d), which is equivalent to Fig. 2 (c) for $\alpha = 1$, $i.e.$, all grounds are shifted by +$V_{S}$/2, which we call a pseudosymmetric bias condition. Hence, when we sweep $V_{S}$, the voltages applied to the gate and substrate electrodes should be simultaneously changed. Note that the pseudosymmetric bias condition is equivalent to the symmetric bias condition only for negligible capacitances between the dominant dot and any other electrodes and/or the ground, except for the gate, substrate, source, and drain electrodes. This is the conventional situation for a trap site in a MOSFET channel, where the capacitances to the gate and substrate dominate the other capacitances.

Under both the symmetric and pseudosymmetric bias conditions, the MOSFET channel is symmetrically biased with source and drain electrodes. However, this does not necessarily mean that the dominant QD itself is symmetrically biased owing to the series resistances in Fig. 1 (b). We first consider the case for negligible series resistances and a symmetrically biased dominant QD.  Suppose we measure the Coulomb diamond data under the (pseudo-) symmetric bias condition and name it data 1. Then, we swap the measurement electronics that is connected to the source electrode with that connected to the drain electrode, and measure the Coulomb diamond again (data 2). Obviously, data 2 should be identical to data 1 except for the sign of $V_{SD}$ ($V_{S}$).

In the case of non-negligible series resistances, Coulomb diamond data 1 and 2 will not be identical because the effective bias voltage for the dominant QD is attenuated by a factor of $\alpha$. Assuming that $\alpha$ is constant for the parameter ranges of interest, this attenuation can be accounted for by replacing $V_{G}$ + $V_{S}$/2 and $V_{Sub}$ + $V_{Sub}$/2 with $V_{G}$ + $\alpha V_{S}$/2 and $V_{Sub}$ + $\alpha V_{S}$/2, respectively. In other words, $\alpha$ can be correctly estimated by finding identical sets of data 1 and 2 obtained for an appropriate $\alpha$. Note that this scheme crucially depends on the series-resistor approximation for the stray QDs, $i.e.$, these series resistors are assumed to not have any capacitances to the dominant dot and/or surrounding electrodes. This approximation should be modified for the stronger charging effect of the stray QDs at lower temperatures and/or weaker tunnel couplings for the stray QDs. 

\begin{figure}
\includegraphics[width=0.45\textwidth, keepaspectratio]{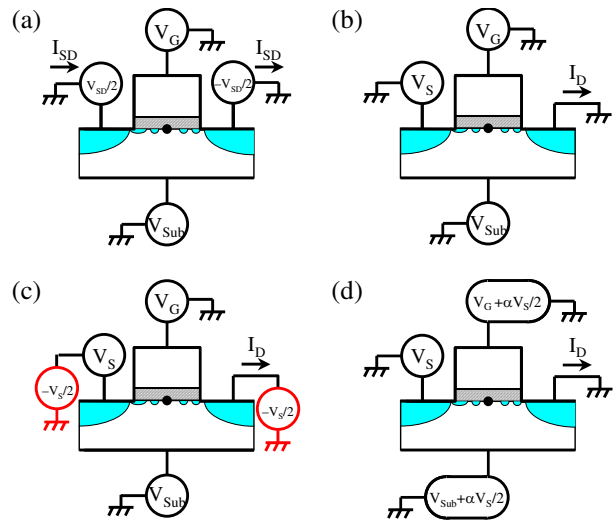}%
\caption{\label{fig:2} 
Measurement conditions for (a) symmetric bias and (b) asymmetric bias. (c) Pseudosymmetric bias condition obtained by replacing the ground for the electronics used to measure $V_{S}$ and $I_{D}$ by -$V_{S}$/2, and (d) condition equivalent to (c) where $V_{G}$ and $V_{Sub}$ are replaced with $V_{G}$ + $V_{S}$/2 and $V_{Sub}$ + $V_{S}$/2, respectively. They can be further replaced with $V_{G}$ + $\alpha V_{S}$/2 and $V_{Sub}$ + $\alpha V_{S}$/2, respectively, to account for the additional voltage drops due to the series resistors.
}
\end{figure}

We measure a p-channel MOSFET with a channel length of 135 nm and a width of 220 nm with a silicon oxynitride gate dielectric, fabricated with standard 130 nm CMOS technologies. The Coulomb diamond measurements are performed in a pumped $^4$He cryostat at temperature $T$ = 1.6 K under the pseudosymmetric bias condition with various $\alpha$ values. We found that some of the MOSFETs exhibit nearly closed Coulomb diamonds in the subthreshold region, indicating that a single dominant QD is present in the channel. In this paper, we focus on the effect of the pseudosymmetric bias on the dominant QD, and the physical origin of the dominant QD will be discussed elsewhere.

\begin{figure*}
\includegraphics[width=0.9\textwidth, keepaspectratio]{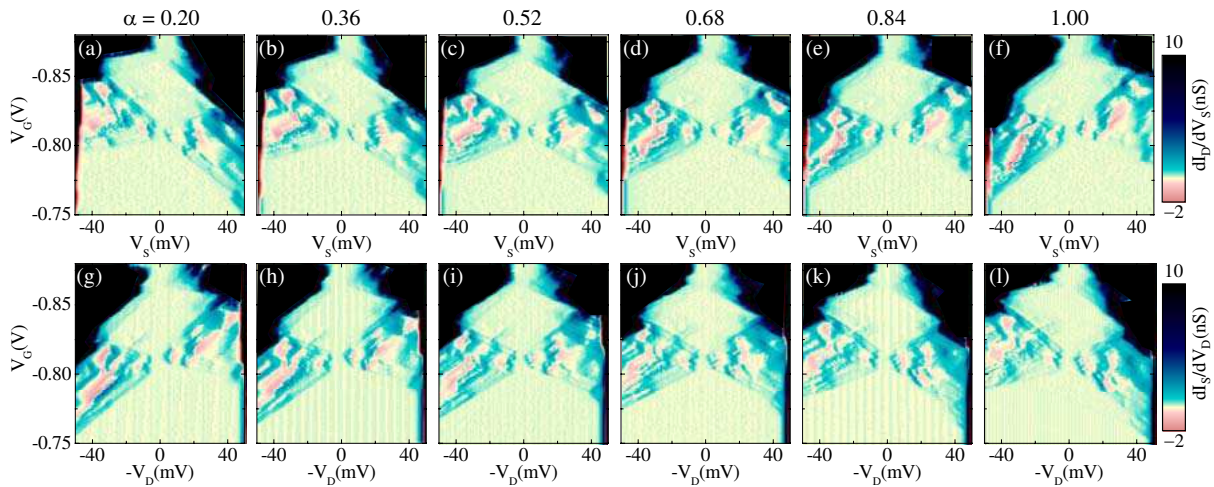}%
\caption{\label{fig:3}
(a)-(f) Intensity plots of differential conductance $dI_{D}/dV_{S}$ as a function of $V_{S}$ and $V_{G}$ under the pseudosymmetric bias condition with (a) $\alpha$ = 0.20, (b) 0.36, (c) 0.52, (d) 0.68, (e) 0.84, and (f) 1.00. (g)-(l) Results of similar measurements with the same $\alpha$ for (a)-(f), respectively, after swapping the electronics for the source and drain electrodes. We apply a drain voltage $V_{D}$ and measure the source current $I_{S}$, applying $V_{G}$ + $\alpha V_{D}$/2 and $V_{Sub}$ + $\alpha V_{D}$/2 to the gate and the substrate electrode, respectively, and plotting $dI_{S}/dV_{D}$ as a function of -$V_{D}$ and $V_{G}$. $V_{Sub}$ is fixed to 0 V throughout these measurements, $i.e.$, $\alpha V_{S}$/2 (or $\alpha V_{D}$/2) is applied to the substrate electrode.
}
\end{figure*}

Figures 3 (a)-(f) show a series of such single-QD-like Coulomb diamonds taken at various $\alpha$ values at zero $V_{Sub}$. A large Coulomb diamond, appearing as the bright region near $V_{G}$ = -0.84 V, of the size in $V_{S}$ = 40 mV is observed. After swapping the source and drain electronics as in the above manner, the data shown in Figs. 3 (g)-(l) are obtained with the same values of $\alpha$ as those in (a)-(f), respectively. The lateral axis for (g)-(l) is reversed for clear identification of the two series of data. It is clearly shown that the two data with $\alpha$ = 0.68 ( (d) and (j)) are nearly identical. Thus, the actual charging energy of the dominant QD defining this Coulomb diamond is estimated to be 40 $\times$ 0.68 = 27 meV. The fact that Figs. 2 (d) and (j) are nearly identical justifies the assumption that $\alpha$ can be regarded as constant for the measured range of $V_{G}$ and  $V_{S}$ (or $V_{D}$).

We found that at 1.6 K, the leakage current of reverse-biased pn junctions between the n-type substrate and the p-type source/drain electrodes is negligibly small ($<$ 1 pA for $V_{Sub}$ = 1.0 V) compared with the typical $I_{D}$ (or $I_{S}$) in the subthreshold region. Thus, applying a positive $V_{Sub}$ essentially has the effect of applying a back gate voltage to the channel, making the subthreshold gate voltage more negative. At room temperature, applying a positive $V_{Sub}$ to p-channel MOSFETs (or a negative $V_{Sub}$ to n-channel MOSFETs) is known to decrease the channel mobility, because the increased effective electric field at the silicon/dielectric interface enhances the electron scattering due to interface roughness \cite{Sun}. Thus, in the subthreshold region with a positive $V_{Sub}$ at a low temperature, it is expected that the potential fluctuation of the channel will increase, thus increasing the charging energies of the stray QDs, and at the same time decreasing the size of the stray QDs.

\begin{figure*}
\includegraphics[width=0.92\textwidth, keepaspectratio]{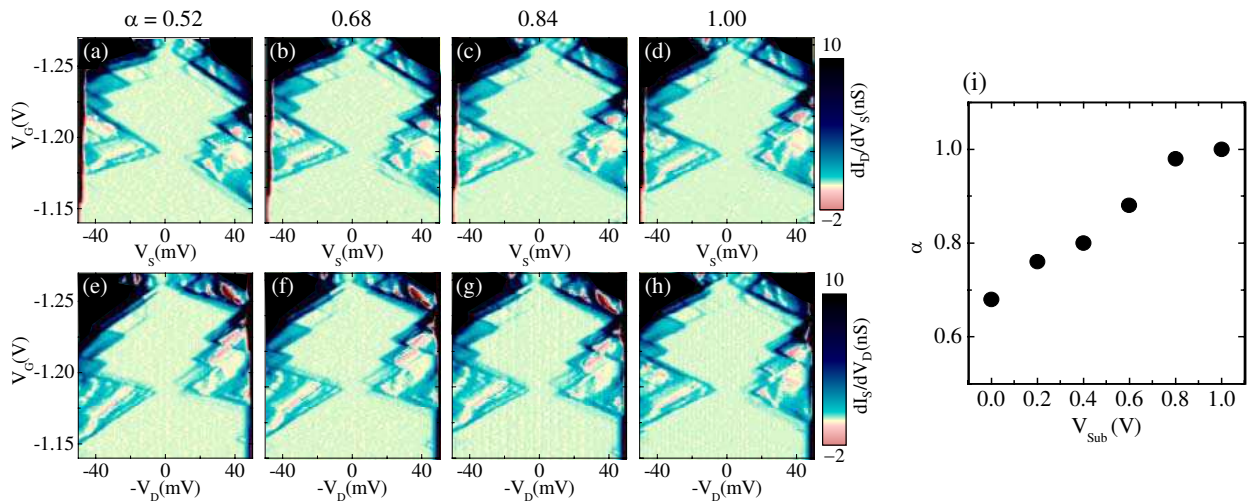}%
\caption{\label{fig:4}
(a)-(d) Intensity plots of differential conductance $dI_{D}/dV_{S}$ for $V_{Sub}$ = 1.0 V with (a) $\alpha$ = 0.52, (b) 0.68, (c) 0.84, and (d) 1.00. (e)-(h) Results after swapping the source and drain electronics in the same manner as in Fig. 3. (i) $V_{Sub}$ dependence of $\alpha$ at which data 1 and 2 becomes most similar. 
}
\end{figure*}

We repeat the measurements shown in Fig. 3 for various positive $V_{Sub}$ values. The results for $V_{Sub}$ = 1.0 V are shown in Figs. 4 (a)-(h). The large Coulomb diamond is shifted to approximately $V_{G}$ = -1.22 V, and its size measured in $V_{S}$ exceeds 40 mV. Data 1 and 2 are nearly identical for $\alpha$ = 1.00 ((d) and (h)). Thus, the size of the diamond is now the correct measure of the charging energy, and the conventional analysis of the Coulomb diamond can be applied for this value of $V_{Sub}$. It was observed that the large diamond is modified by a series of kink structures with a typical width of $\sim$ 5 mV measured in $V_{S}$ (or -$V_{D}$). Moreover, the large diamond is nearly closed but a small opening of $\sim \pm$ 5 mV exists at $V_{G}$ = -1.18 V. These findings indicate, on the basis of the combinational analysis of multiple QDs, that there is at least one QD with a charging energy of $\sim$ 5 meV connected in series to the dominant QD, whose charging energy is $\sim$ 40 meV. 
In ref. \cite{Sellier}, Sellier \emph{et al.} observed single-dot-like transport (two-step tunneling) via a single donor site in a MOSFET with 60 nm channel length. Considering the fact that a stray QD with smaller charging energy has a larger size, the channel length of our device, 135 nm, is not large for such double-dot-like transport (three-step tunneling) between source and drain electrodes via a dominant QD and at least one stray QD. Furthermore, the data and analysis do not exclude the presence of other stray QDs with even smaller charging energies (thus even larger sizes). Such large stray QDs near the source and/or drain electrode have little effect on the Coulomb diamond and essentially behave as part of the electrode and reduce the effective channel length. 

Figure 4 (i) shows a plot of $\alpha$ that gives the most identical data 1 and 2 for each $V_{Sub}$.  $\alpha$ asymptotically approaches 1 with increasing $V_{Sub}$. This is interpreted as follows. With increasing $V_{Sub}$ from 0, some of the stray QDs that have relatively large charging energies will not behave as part of the series resistor but as QDs with a non-negligible charging effect. At a large positive $V_{Sub}$, nearly all the stray QDs exhibit a well-defined charging effect with a minor series resistor effect, where conventional multiple QD analysis can be effectively applied.
In the measurements in Figs. 3 and 4 (a)-(h), the voltage applied to the substrate electrode is changed while sweeping $V_{S}$ (or $V_{D}$). To justify the assumption of a constant $\alpha$ during the sweeping of $V_{S}$, the change in $\alpha$ should be small enough in the scale of half of the amplitude of the $V_{S}$ sweep, 50 mV. As shown in Fig. 4 (i), $\alpha$ indeed slowly changes with $V_{Sub}$ in the scale of $\sim$ 50 mV.

The observed $\alpha$ also should be, in principle, verified by measuring the temperature dependence of the thermal-activation-type conductance at zero-bias voltage,

\begin{equation}
\frac{dI_{D}}{dV_{S}} = G_{0} \exp (-\frac{E_{a}}{2k_{B}T}),
\end{equation}

\noindent where $E_{a}$ is the activation energy, $G_{0}$ is the characteristic conductance, and $k_{B}$ is the Boltzmann constant [1]. $E_{a}$ corresponds to the charging energy at the center of the Coulomb diamond. However, we show that this is not the case for the dominant QD in the MOSFET channel. We measured the zero-bias conductance in the temperature range from 1.6 K to 130 K at $V_{Sub}$ = 0 V. Figure 1 (a) shows an activation plot of the zero-bias conductance at $V_{G}$ = -0.842 V, where the Coulomb diamond takes its maximum width in $V_{S}$. Similar activation-type temperature dependence is observed at various $V_{G}$ values around the Coulomb diamond as plotted in Fig. 1 (b). If the activation-type conductance is limited by the dominant QD, $E_{a}$ should decrease to zero at the values of $V_{G}$ where the Coulomb diamond is nearly closed. However, this is not the case. Figure 1 (b) shows a monotonic change in $E_{a}$ with $V_{G}$, regardless of the Coulomb diamond. Moreover, the value of $G_{0}$ obtained from the activation fit of the data in Fig. 1 (a) is 4.6 $\mu$S, which is very large compared with the typical differential conductance of order 1 nS around the Coulomb diamond (see Fig. 3 of main manuscript). Values of $G_{0}$ of similar order are obtained at other $V_{G}$ values for the data in Fig. 1 (b). All these findings indicate that the observed $E_{a}$ is not due to the charging energy of the dominant QD but the thermally activated transport of the closed channel itself. Indeed $E_{a}$ appears to decrease for more negative $V_{G}$ and vanish around the gate threshold of the channel, $\sim$ -0.9 V, at which the device shows a typical zero-bias conductance of order 1 $\mu$S. The temperature dependence of the zero-bias conductance for the dominant QD, which should have a similar $E_{a}$ but a much smaller $G_{0}$ of order 1 nS, is masked by this thermally activated channel conductance, and thus the measured temperature dependence is not helpful for determining the charging energy of the dominant QD in the sub-threshold region.

Note that we have measured several p-MOSFETs with similar gate lengths, and some of them exhibited large "unclosed" Coulomb diamonds with a large opening in $V_{S}$, indicating that two or more dominant QDs exist in the channel. Such a channel can be regarded as a system with multiple dominant QDs and possible series resistors. By applying pseudosymmetric bias to such a system, we can at least confirm the existence of non-negligible stray resistors and determine whether or not the conventional analysis of the Coulomb diamond for multiple dots is applicable. However estimation of each attenuation factors for each dominant QD is a difficult problem and is beyond the scope of this paper.

 We described the pseudosymmetric bias method for correctly estimating the charging energy of the single dominant QD in a MOSFET channel. The result for the p-channel MOSFET shows that the charging energy of the dominant QD can be 0.68 times the value measured from the size of the Coulomb diamond in $V_{S}$ because the array of stray QDs with a weak charging effect causes additional drops of $V_{S}$. Applying a positive $V_{Sub}$ increases the charging effect of the stray QDs and the conventional multiple QD analysis can be applied with a minor series resistor effect.

We would like to thank M. Kawamura, S. Amaha, K. Kono, K. Kato, and K. Itoh for various discussions and help.

\begin{figure}
\includegraphics[width=0.45\textwidth, keepaspectratio]{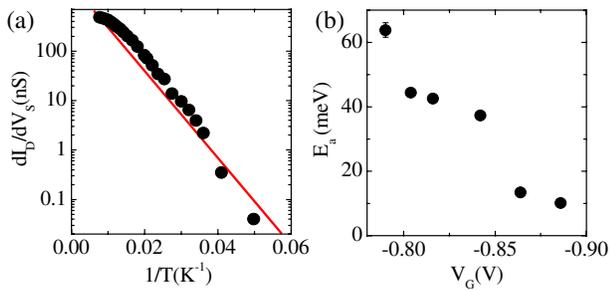}%
\caption{
(a) Temperature dependence of zero-bias conductance measured at $V_{S}$ = 0.1 mV, $V_{G}$ = -0.842 V, and $V_{Sub}$ = 0 V. (b) $V_{G}$ dependence of the activation energy $E_{a}$ obtained by fitting with Eq. (1). 
}
\end{figure}


\end{document}